\documentclass[a4paper,fleqn,usenatbib]{mnras}
\usepackage{txfonts}

\usepackage[T1]{fontenc}
\usepackage{ae,aecompl}

\usepackage{xspace}
\usepackage{etex}
\usepackage{graphics,graphicx}
\usepackage{natbib}
\usepackage{multirow}
\usepackage[percent]{overpic}
\usepackage{color}

\newcommand{\as}{\mbox{\ensuremath{.\!\!^{\prime\prime}}}}

\newcommand{\asn}{$^{\prime\prime}$\xspace}

\newcommand{\nH}{$N_{\rm H}$\xspace}
\newcommand{\PL}{$\Gamma$\xspace}
\newcommand{\Msun}{$M_{\odot}$\xspace}

\newcommand{\chisq}{$\chi^2$\xspace}

\newcommand{\Chandra}{{\it Chandra}\xspace}
\newcommand{\HST}{{\it HST}\xspace}
\newcommand{\Swift}{{\it Swift}\xspace}
\newcommand{\XMM}{{\it XMM-Newton}\xspace}
\newcommand{\lum}{erg s$^{-1}$\xspace}
\newcommand{\flux}{erg s$^{-1}$ cm$^{-2}$\xspace}

\title[X-ray Outbursts in SN~2010da]{Recurring X-ray Outbursts in the Supernova Impostor SN~2010da in NGC~300}

\author[Binder et al.]{
B. Binder$^{1}$\thanks{E-mail: bbinder@uw.edu}, 
B. F. Williams$^{1}$, 
A. K. H. Kong$^{2}$, 
T. J. Gaetz$^{3}$, 
P. P. Plucinsky$^{3}$, 
\newauthor E. D. Skillman$^{4}$, 
A. Dolphin$^{5}$	\\
$^{1}$University of Washington, Department of Astronomy, Box 351580, Seattle, WA, 98195 USA \\
$^{2}$Institute of Astronomy and Department of Physics, National Tsing Hua University, Hsinchu 30013, Taiwan \\
$^{3}$Harvard-Smithsonian Center for Astrophysics, Cambridge, MA 02138, USA \\
$^{4}$Astronomy Department, University of Minnesota, 116 Church St. SE, Minneapolis, MN 55455, USA \\
$^{5}$Raytheon Company, Tucson, AZ 85734, USA 
}

\date{Accepted XXX. Received YYY; in original form ZZZ}
\pubyear{2016}

\begin{document}
\label{firstpage}
\pagerange{\pageref{firstpage}--\pageref{lastpage}}
\maketitle

\begin{abstract}
We present new observations of the ``supernova impostor" SN~2010da using the \Chandra X-ray Observatory and the {\it Hubble Space Telescope}. During the initial 2010 outburst, the 0.3-10 keV luminosity was observed by {\it Swift} to be $\sim5\times10^{38}$ \lum and faded by a factor of $\sim$25 in a four month period. Our two new \Chandra observations show a factor of $\sim$10 increase in the 0.35-8 keV X-ray flux, from $\sim$4$\times10^{36}$ \lum to $4\times10^{37}$ \lum in $\sim$6 months, and the X-ray spectrum is consistent in both observations with a power law photon index of \PL$\sim0$.  We find evidence of X-ray spectral state changes: when SN~2010da is in a high-luminosity state, the X-ray spectrum is harder (\PL$\sim0$) compared to the low-luminosity state (\PL$\sim1.2\pm0.8$). Using our {\it Hubble} observations, we fit the color magnitude diagram of the coeval stellar population to estimate a time since formation of the SN~2010da progenitor system of $\lesssim$5 Myr. Our observations are consistent with SN~2010da being a high-mass X-ray binary (HMXB) composed of a neutron star and a luminous blue variable-like companion, although we cannot rule out the possibility that SN~2010da is an unusually X-ray bright massive star. The $\lesssim$5 Myr age is consistent with the theoretically predicted delay time between the formation of a massive binary and the onset of the HMXB phase. It is possible that the initial 2010 outburst marked the beginning of X-ray production in the system, making SN~2010da possibly the first massive progenitor binary ever observed to evolve into an HMXB.
\end{abstract}



\section{Introduction}
On 2010 May 24, an optical transient ($M_V\sim-10.3$) was detected in the nearby spiral galaxy NGC~300 \citep{Monard10} and given the supernova (SN) designation SN~2010da. Subsequent observations revealed that the event was not a true SN, but a supernova impostor\footnote{The term SN ``impostor'' was first coined by \cite{vanDyk+00}, although the category was initially referred to as ``Type~V'' SNe by \cite{Zwicky64}.}: optical spectroscopy revealed that the transient was more similar to a luminous blue variable (LBV)-like outburst from a dust enshrouded massive star \citep{Chornock+10,EliasRosa+10a,EliasRosa+10b}. The outburst was additionally observed by the {\it Swift} X-ray Telescope, which found an X-ray source with a 0.2-10 keV luminosity of $\sim5\times10^{38}$ \lum coincident with the position of the optical outburst \citep{Immler+10}. Four months after the outburst, SN~2010da was observed with \Chandra and found to have decreased in luminosity by a factor of $\sim$25 \citep{Binder+11}. 

Prior to the initial outburst, no X-ray emission from the location of SN~2010da had been detected, despite having been observed four times by \XMM. The progenitor optical counterpart was not detected in archival Magellan/Megacam images to a limiting magnitude of 24 AB mag \citep{Berger+10}. The progenitor was, however, detected in the mid-IR by {\it Spitzer} and found to be consistent with an LBV-like star enshrouded in dust \citep{Khan+10}; the {\it Swift} UVOT observations indicated that some of the dust was destroyed during the outburst \citep{Brown10}, although follow-up observations by \cite{Prieto+10} are consistent with some of the circumstellar dust having survived the event. Archival IR observations with {\it Spitzer} additionally showed a significant brightening of the progenitor system $\sim$6 months prior to the outburst \citep{Laskar+10}. Follow-up spectroscopy with GMOS on Gemini-South reported by \cite{Chornock+10} in an Astronomer's Telegram revealed H$\alpha$ emission (with a FWHM of $\sim$660 km s$^{-1}$), as well as emission lines of He~I, Fe~II, Ca~II K and H, O~I, and He~II $\lambda$4648, with no strong absorption from Na~I D seen. However, the spectrum was not formally published, and no other line velocities have been reported. Monitoring of this system with the Magellan/Clay 6.5-m telescope by \cite{Chornock+11} revealed a re-brightening of the optical counterpart from $i\sim$20.7 on 2011 January 13 to $i\sim18.7$ on 2011 October 21. This second re-brightening event was also followed up by {\it Swift}, but no accompanying X-ray emission was detected \citep[down to a 0.3-10 keV limiting luminosity of $\sim10^{37}$ \lum;][]{Chornock+11}.

There are three main scenarios that can explain X-ray emission from massive stars: intrinsic X-ray emission, colliding winds in binary systems, and accretion onto compact objects. The intrinsic X-ray emission from early type stars is typically thermal in origin with $kT\sim0.5$ keV and X-ray luminosities of $\sim10^{31-33}$ \lum \citep{Lucy+80,Wang+07,Skinner+10}. Harder X-ray emission can be produced in a massive binary, where the winds generated from two stars collide and with luminosities in the range of $\sim10^{32-34}$ \lum \citep{Mauerhan+10}. Neither intrinsic X-ray emission nor emission arising from a colliding wind has been observed to exceed luminosities of $\sim10^{35}$ \lum \citep[see ][their Figure~6]{Guerrero+08}, and no known LBV has been observed to have an X-ray luminosity above $\sim$10$^{34}$ \lum \citep{Naze+12}, although the X-ray properties during and immediately following an eruption are largely unknown. The high X-ray luminosities, even when not in optical outburst ($\gtrsim$2 orders of magnitude higher than even the most luminous colliding wind binaries or single stars), could be explained by the presence of a neutron star (NS) in a binary system as was suggested in \cite{Binder+11}, and the presence of the He~II $\lambda$4686 emission line in the optical spectrum of SN~2010da \citep{Chornock+10,EliasRosa+10b,Chornock+11} is consistent with hotspots in a NS accretion disk \citep{Still+97,ValBaker+05,Pearson+06}. 

Although the high X-ray luminosity of SN~2010da is consistent with outbursting high mass X-ray binary (HMXB) with a NS primary and an LBV-like companion \citep{Binder+11}, many details of the system are not yet known. Roughly 60\% of known HMXBs contain fast-rotating Be stars and a NS in a wide, moderately eccentric orbit \citep[e.g, BeXBs,][]{Liu+06}, and analysis of the coeval stellar populations have revealed a preferred age of $\sim$40-55 Myr \citep{Antoniou+09,Williams+13}. In these systems, X-ray outbursts reaching $\sim10^{37}$ \lum are observed to correspond with the periastron passage of the NS (e.g, ``Type I'' outbursts), and even more luminous ``Type II'' outbursts are observed to last over several orbital periods \citep{Reig08}. HMXBs containing supergiant OB companions are less numerous, and most are thought to undergo bright ($\sim10^{36-37}$ \lum) and short X-ray flares lasting from minutes to hours \citep[see][and references therein]{Ducci+14}. {\it Swift} monitoring of supergiant fast X-ray transients (SFXTs) has revealed that shorter X-ray flares are frequently part of much longer outburst events, which may last for several days \citep{Romano+07, Sidoli+08, Romano+08}. The mechanism by which these flares are produced is not certain, although eruptions from the donor star or changes in the stellar wind properties may contribute to the observed X-ray variability \citep[][and references therein]{Ducci+14}. 

We have obtained two new \Chandra observations of SN~2010da, as well as four orbits on the {\it Hubble Space Telescope} (\HST). This is the first time the SN~2010da region has been imaged with \HST, and SN~2010da is clearly detected in both new \Chandra observations. By analyzing the resolved stellar population in the vicinity of SN~2010da, we are able to put limits on the age of the system. Throughout this work, we assume a distance to NGC~300 of 2.0 Mpc \citep{Dalcanton+09}. Our observations and data reduction procedures are described in Section~\ref{section_observations}. In Section~\ref{section_xray} we present an analysis of the X-ray data, including imaging, spectral fitting, and variability measurements. In Section~\ref{section_optical} we present optical imaging of the companion star and constrain the age of the SN~2010da system. We conclude with a discussion of our findings in Section~\ref{section_discussion}.


\begin{table*}
\setlength{\tabcolsep}{4pt}
\centering
\caption{\Chandra Observation Log \& Alignment to USNO-B1.0}
\begin{tabular}{cccccccc}
\hline \hline
\multirow{2}{*}{Obs. ID} & \multirow{2}{*}{Date}	& Exposure	& \# Sources Used	& rms Residuals	& \multicolumn{2}{c}{SN~2010da Position}	& Net Counts	\\
			&						& Time (ks)	& for Alignment		& (arcsec)			& R.A. (J2000)		& Decl. (J2000)			& (0.35-8 keV)	\\
 (1)			& (2)						& (3)      		& (4)	 			& (5) 			& (6)				& (7)					& (8)			\\
\hline
12238		& 2010 Sept. 24	& 63.0		& 7		& 0\as267		& 00:55:04.87	& -37:41:44.1	& 77$\pm$9	\\
16028		& 2014 May 16-17	& 63.9		& 5		& 0\as429		& 00:55:04.93	& -37:41:43.9	& 25$\pm$5 	\\
16029		& 2014 Nov. 17-18	& 61.3		& 9 		& 0\as423		& 00:55:04.83	& -37:41:43.6	& 137$\pm$12	\\
\hline \hline
\end{tabular}
\label{table_log}
\end{table*}

\begin{figure}
\centering
\begin{tabular}{cc}
\includegraphics[width=1\linewidth,clip=true,trim=0cm 0cm 0cm 0cm]{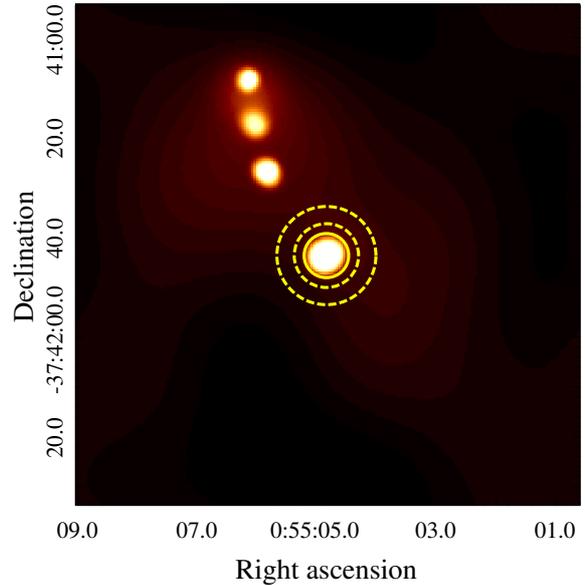} 
\end{tabular}
\caption{The adaptively smoothed 0.35-8 keV image of the SN~2010da region, using all three \Chandra observations in a combined image. The inner yellow circle shows the source extraction region, while the dashed annulus shows the background extraction region. Three additional sources are also visible in the combined image.}
\label{figure_xray_imaging}
\end{figure}

\section{Observations and Data Reduction}\label{section_observations}
	\subsection{X-ray Observations from \Chandra}
Two new $\sim$65 ks observations of the nearby Sculptor group galaxy NGC~300 were obtained with the \Chandra Advanced CCD Imaging Spectrometer (ACIS-I) on 2014 May 16-17 and 2014 November 17-18. An additional $\sim$65 ks observation with ACIS-I from 2010 is additionally available in the literature \citep{Binder+11}. All three observations contained SN~2010da within the field of view. All three exposures were reprocessed from the \texttt{evt1} level files with the  \Chandra Interactive Analysis of Observations (CIAO) software package version 4.6.1 and CALDB version 4.6.3 using the task \texttt{chandra\_repro}. We inspected background light curves for all three observations and did not find any significant flaring events. The task \texttt{lc\_clean} and 5$\sigma$ clipping was used to generate good time intervals (GTIs) for each observation, and all event data were filtered on these GTIs. 

We performed relative astrometry to align the \Chandra \texttt{evt2} data for all three observations to the USNO-B1.0 catalog using \texttt{wcs\_match} and \texttt{wcs\_update} in CIAO, and \texttt{fluximage} was used to construct exposure maps for all three exposures. We produced exposure-corrected images using a custom instrument map, where we assume a power-law spectrum with \PL=1.7 absorbed by the average foreground column density \nH = 4.09$\times10^{20}$ cm$^{-2}$ \citep[][appropriate for both XRBs and AGN]{Kalberla+05}. The \texttt{wavdetect} task was used to detect point sources, and sources with a signal-to-noise ratio $\geq3$ were considered significant.

Table~\ref{table_log} provides an observation log of all three observations of SN~2010da, including: the observation ID number (hereafter referred to as the ``ObsID''), the date of the observation, the effective exposure time, number of sources used for our relative astrometry, the root-mean-square (rms) residuals for our alignment after aligning to USNO-B1.0, the \texttt{wavdetect} position of the X-ray point source (RA and Decl.), and the number of 0.35-8 keV net counts detected in each ObsID. SN~2010da was detected at high significance in all three observations, at positions in excellent agreement with the position reported in SIMBAD \citep{Wegner+00}. 

Figure~\ref{figure_xray_imaging} shows the exposure-corrected, adaptively smoothed 0.35-8 keV \Chandra image of SN~2010da, generated using the CIAO task \texttt{csmooth}, from the combined image from all three observations (hereafter referred to as the ``merged'' image). To determine source and background extraction regions, we extracted the 0.35-8 keV radial surface brightness profile centered on the position of SN~2010da. The circular source extraction region was defined by the radius that enclosed $\sim$90\% of the \Chandra point spread function (PSF) at the source position. The background region was a circular annulus with the inner radius set at the distance where the surface brightness profile reached the background level. The outer radius was determined such that the source and background regions contained roughly equal numbers of pixels. Source and background regions were visually inspected to ensure that no obvious spurious sources were contained within either region. Three fainter X-ray sources were detected in the merged image, but were outside our source and background extraction regions.

	\subsection{Optical Observations from \HST}\label{HST_imaging}
In addition to our two new \Chandra exposures, we obtained \HST imaging to study the resolved stellar populations in the immediate vicinity of SN~2010da. The observations were taken with ACS on 2014 July 2 using the $F606W$ and $F814W$ filters and on 2014 September 5 using the UVIS $F336W$ filter.

\begin{figure}
\centering
\includegraphics[width=0.75\linewidth,clip=true,trim=0cm 0cm 0cm 0cm]{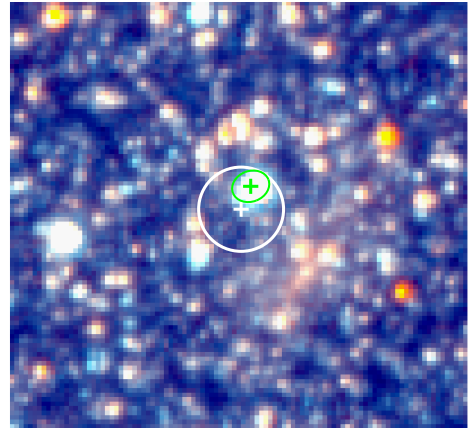}
\caption{The RGB rendered \HST image of the SN~2010da optical counterpart. The red channel contains the $F814W$ image, the green channel contains the $F606W$ image, and the blue channel was estimated as $2\times F606W-F814W$. The X-ray error circle is shown by the white circle (with a radius of 0\as47), and the white cross indicates the X-ray position. The green cross and circle show the position and uncertainty of the optical counterpart after aligning to the USNO-B1.0 catalog. The image dimensions are 5\asn $\times$ 5\asn (where 5\asn$\approx50$ pc at the distance of NGC~300).}
\label{figure_HST_aligned}
\end{figure}

 To measure resolved stellar photometry in our \HST images, we apply the same technique as used for the ACS imaging of the Panchromatic Hubble Andromeda Treasury \citep{Williams+14}. Briefly, we used the point spread function photometry package DOLPHOT \citep[an updated version of HSTPHOT;][]{Dolphin00}. In DOLPHOT, all of the individual CCD exposures are aligned and stacked in memory to search for any significant peaks, and then each significant peak above the background level is fitted with the appropriate point spread function. The measurements are corrected for charge transfer efficiency and calibrated to infinite aperture. The measurements are then combined into a final measurement of the photometry of the star, and include a combined value for all data in each band for the count rate, rate error, VEGA magnitude and error, background, $\chi^2$ of the PSF fit, sharpness, roundness, crowding, and signal-to-noise.  For the photometry reported here, we use the combined VEGA magnitude in each observed band.

To directly compare our X-ray observations with the optical ones, we first needed to place both sets of observations on the same coordinate system. Directly aligning both the \Chandra and \HST images was not reliable due to the difficulty matching X-ray sources with optical counterparts in the \HST field. We therefore aligned both the X-ray and optical images to the same large-field optical reference image and coordinate system. To do this, we retrieved a publicly-available $B$-band image from NED\footnote{\url{http://ned.ipac.caltech.edu/}} \citep{Larsen+99}. The reference $B$-band image was aligned using the USNO-B1.0 catalog using the positions of six stars. The IRAF task \texttt{ccmap} was used to compute the plate solution and to update the image header with corrected WCS information; the root-mean square (rms) residuals of the fit were 0\as16 in both right ascension and in declination. We next aligned our \HST fields to the ground-based $B$-band image by identifying five matched sources with the USNO-B1.0 catalog. The resulting positional uncertainty in our \HST field is 0\as21 in right ascension and 0\as17 in declination.

Figure~\ref{figure_HST_aligned} shows a 5\asn $\times$ 5\asn, RGB rendered image of the optical counterpart. To create an RGB-rendered image from our \HST observations, the $F814W$ image is set to the red channel, the $F606W$ filter is used for the green channel, and the blue channel is estimated by taking $2\times F606W-F814W$. The X-ray error circle is shown in white and has a radius of 0\as47. The optical counterpart is detected at R.A. (J2000) 00:55:0.482 and Decl. (J2000) -37:41:43.0, with $F606W$ and $F814W$ magnitudes of $m_{606}=20.52\pm0.01$ and $m_{814}=20.52\pm0.01$. This is comparable to the $i\sim20.7$ AB mag measured in 2011 January \citep[ATel \#3726;][]{Chornock+11}, before the star exhibited a re-brightening of $\sim$2 mag by October of the same year \citep[ATels \#2636 and 2637][]{Chornock+10,EliasRosa+10a,EliasRosa+10b}.

\section{Analysis of the \Chandra X-ray Observations}\label{section_xray}
Directly fitting the X-ray spectrum of SN~2010da is not feasible for the ObsID 16028 observation, as only $\sim$25 net counts were detected in the 0.35-8 keV band. We extract spectra and response files for the ObsID 16029 observation (containing $\sim$140 net counts) in the 0.35-8 keV band using \texttt{specextract}, and \texttt{XSPEC} \citep{Arnaud96} v.12.6.0q was used to perform all spectral fitting. The spectrum was fit using two methods: binning the spectrum to contain 10 counts per bin and using standard \chisq statistics, and fitting the unbinned spectrum using $C$-statistics \citep{Cash79}. When using $C$-statistics, the \texttt{XSPEC} command ``goodness'' was used to generate $10^4$ Monte Carlo realizations of the SN~2010da spectrum using the current best-fit model and return the percentage of simulated spectra that had a fit statistic less than that obtained from the fit to the observed data. Values near 50\% are indicative of a model that well describes the observed data (very low percentages indicate that the data is over-parameterized by the assumed model, while very high percentages indicate the model is a poor fit to the data). Both methods yielded similar fit parameters; we report the best-fit parameters and goodness of fit statistic obtained by the \chisq fitting method.
 
\begin{figure*}
\centering
\includegraphics[width=0.45\linewidth,angle=-90,clip=true,trim=1.5cm 0cm 0cm 0cm]{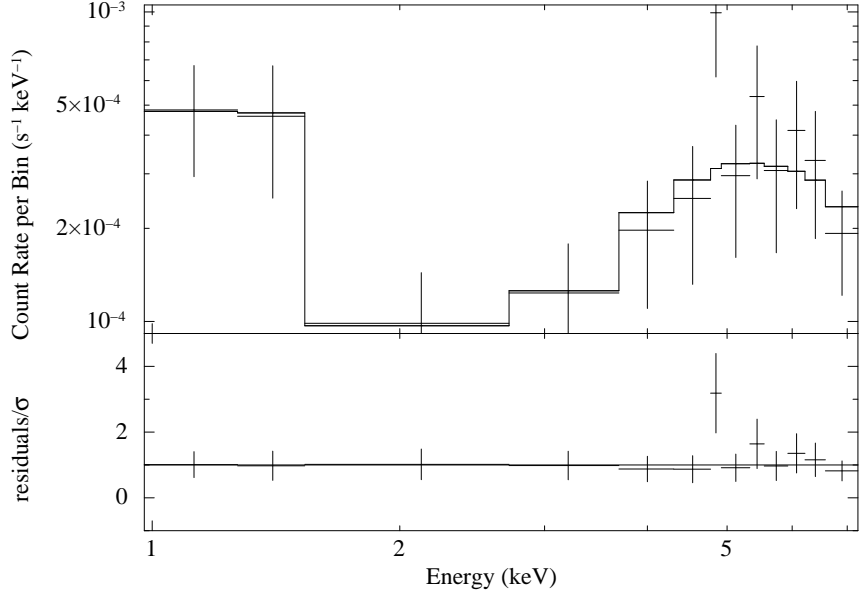}
\caption{The 0.35-8 keV spectrum of SN~2010da from ObsID 16029, binned to 10 counts per bin, with the best-fit bremsstrahlung and power law model superimposed.}
\label{figure_spectrum}
\end{figure*}

The ObsID 16029 spectrum was found to be hard, similar to what was observed in ObsID 12238 \citep{Binder+11} and during the initial outburst with {\it Swift} \citep{Immler+10}. We first fit the spectrum with two simple models: a power law and a blackbody, both with a column of neutral absorption fixed at the Galactic column \citep[\nH = 4.09$\times10^{20}$ cm$^{-2}$;][]{Kalberla+05}. The power law model was consistent with a photon index of \PL$=0.0\pm0.4$ (\chisq/dof = 16/11), and the simple blackbody model yielded a best-fit temperature of $kT=2.6^{+1.7}_{-0.7}$ keV and a \chisq/dof = 18/11. Allowing the absorbing column to vary did not improve the quality of either fit. 

In addition to significant hard X-ray emission near $\sim$6-8 keV, the spectrum shows an excess of soft emission near $\sim$0.7-1 keV. We therefore added a soft blackbody component to our simple power law (\PL$\sim$0) model. We find $kT=0.2^{+0.2}_{-0.1}$ keV and \chisq/dof = 5/8, and find an upper limit on the absorbing column of $7\times10^{21}$ cm$^{-2}$. Nearly identical fit parameters are found when we replace the blackbody component with a thermal bremsstrahlung component. We used the \texttt{XSPEC} routine \texttt{ftest} to compare the power law-only model to the power law plus thermal component model, and found an $F$-probability of $\sim$2\%. We therefore find marginal evidence of both thermal and hard X-ray emission in the spectrum of SN~2010da. The predicted model (unabsorbed) 0.35-8 keV flux is $(8.4\pm3.6)\times10^{-14}$ \flux, corresponding to an unabsorbed luminosity of $(4.0\pm1.6)\times10^{37}$ \lum at the distance of NGC~300 (comparable to what was observed in ObsID 12238), with $\gtrsim$90\% of the X-ray flux originating in the power law component. The 0.35-8 keV spectrum of SN~2010da, with the best-fit model superimposed, is shown in Figure~\ref{figure_spectrum}.

Although we cannot directly model the X-ray spectrum of ObsID 16028, we can use X-ray hardness ratios to search for evidence of spectral variability as a function of the observed flux. We define a hardness ratio as $(\mathcal{H}-\mathcal{M})/(\mathcal{H}+\mathcal{M})$, where $\mathcal{M}$ is defined as the medium energy 1-2 keV band and $\mathcal{H}$ is defined as the hard 2-8 keV band. These energy bands were chosen because the SN~2010da count rate in the soft 0.35-1 keV band was low in all three observations, leading to large uncertainties in the hardness ratio when the soft band was used. Using \texttt{WebPIMMS}, we estimated the \Chandra ACIS-I count rates in the $\mathcal{M}$ and $\mathcal{H}$ bands for a source with power law indices ranging from \PL$=0-2.5$ (as our spectral fitting indicates that absorption intrinsic to the source is low, we fix \nH in our calculations to the Galactic column) and computed the corresponding hardness ratios. Our calculations are summarized in Table~\ref{table_predicted_PL}.

\begin{table}
\centering
\caption{Predicted Hardness Ratios for Power Law Photon Indices}
\begin{tabular}{cc}
\hline \hline
Power Law \PL		& (H-M)/(H+M)		\\
(1)				& (2)				\\
\hline
0.0	& 0.59	\\
0.2	& 0.54	\\
0.4	& 0.46	\\	
0.5	& 0.42	\\
0.8	& 0.29	\\
1.0	& 0.20	\\
1.2	& 0.12	\\
1.5	& 0.00	\\
2.0	& -0.24	\\
2.5	& -0.43	\\
\hline \hline
\end{tabular}
\label{table_predicted_PL}
\end{table}

Our measured hardness ratios for SN~2010da using the definition above are 0.56$^{+0.09}_{-0.10}$ in ObsID 12238, 0.12$^{+0.34}_{-0.37}$ in ObsID 16028, and 0.60$^{+0.06}_{-0.07}$ in ObsID 16029. The uncertainties were estimated from the (assumed Poisson) uncertainties in the count rate in each energy band. The hardness ratios in ObsID 12238 and 16029, when the 0.35-8 keV X-ray luminosity was a few $10^{37}$ \lum, are similar, while the hardness ratio measured when SN~2010da was fainter is consistent with a slightly softer spectrum. Using the predicted values of the hardness ratio, we infer the power law photon index during ObsID 16028 to be \PL$=1.2\pm0.8$. The inferred values of \PL for the two more luminous detections are consistent with the results of our spectral modeling. Using the observed 0.35-8 keV count rate and the inferred value of \PL for ObsID 16028, we estimate the corresponding unabsorbed luminosity during this observation to be $\sim4\times10^{36}$ \lum.

\begin{figure}
\centering
\includegraphics[width=1\linewidth,clip=true,trim=0.5cm 0.4cm 0.2cm 0.2cm]{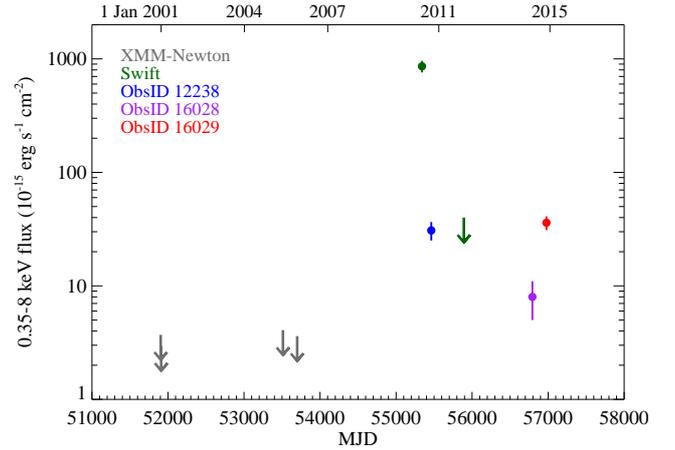}
\caption{The long-term light curve of SN~2010da from \XMM (gray), \Swift (green), and \Chandra. ObsID 12238 is shown in blue, ObsID 16028 is shown in purple, and ObsID 16029 is shown in red. Upper limits correspond to 3$\sigma$ significance. }
\label{figure_light_curve}
\end{figure}

Figure~\ref{figure_light_curve} shows the long-term luminosity evolution of SN~2010da, constructed using all available \XMM, {\it Swift}, and \Chandra data. Prior to the optical outburst that led to this source's discovery, it was undetected in four \XMM exposures; the 3$\sigma$ upper limits were estimated using the Flux Limits from Images from \XMM (\texttt{FLIX})\footnote{\url{http://www.ledas.ac.uk/flix/flix.html}}. The more recent {\it Swift} observation, taken in 2011 following the observed re-brightening of the optical companion, also failed to detect SN~2010da, although the exposure was significantly shallower than the \Chandra observations. The upper limits in Figure~\ref{figure_light_curve} represent 3$\sigma$ significance. To search for rapid variability in SN~2010da, we extracted 0.35-8 keV light curves from all three \Chandra observations using the CIAO tool \texttt{dmextract}. The light curves were analyzed using the Gregory-Loredo variability program \texttt{glvary} in CIAO, and no evidence of rapid variability was found in any of the observations (the probability of the light curves being variable was $<$17\% in all observations). The lack of rapid variability is likely due to the low number of counts observed in each observation. In ObsID 16029, the X-ray emission from SN~2010da would have needed to vary by a factor of $\gtrsim$20 to be detected at 95\% significance. We estimate that in order to observe variability with a factor of $\geq$5 change over a $\sim$few ks at 95\% confidence would require $\gtrsim$400 counts.

\section{The Age of SN~2010da}\label{section_optical}
Most stars form in stellar clusters, which share a common age and metallicity \citep{Lada+03}, and these stars remain spatially correlated on scales of up to $\sim$100 pc for $\sim$100 Myr after their formation \citep{Bastian+06}. The age of an individual HMXB can therefore be recovered from the color-magnitude diagram (CMD) of the surrounding coeval stellar population. \cite{Williams+13} used this technique to measure the ages of 18 HMXB candidates in NGC~300 and NGC~2403, and found a typical HMXB age of $\sim$40-55 Myr. This same approach was also used to constrain the mass of the 2008 optical transient in NGC~300 \citep{Gogarten+09}.

\begin{figure*}
\centering
\begin{tabular}{cc}
\includegraphics[width=0.5\linewidth,clip=true,trim=0cm 0cm 0cm 0cm]{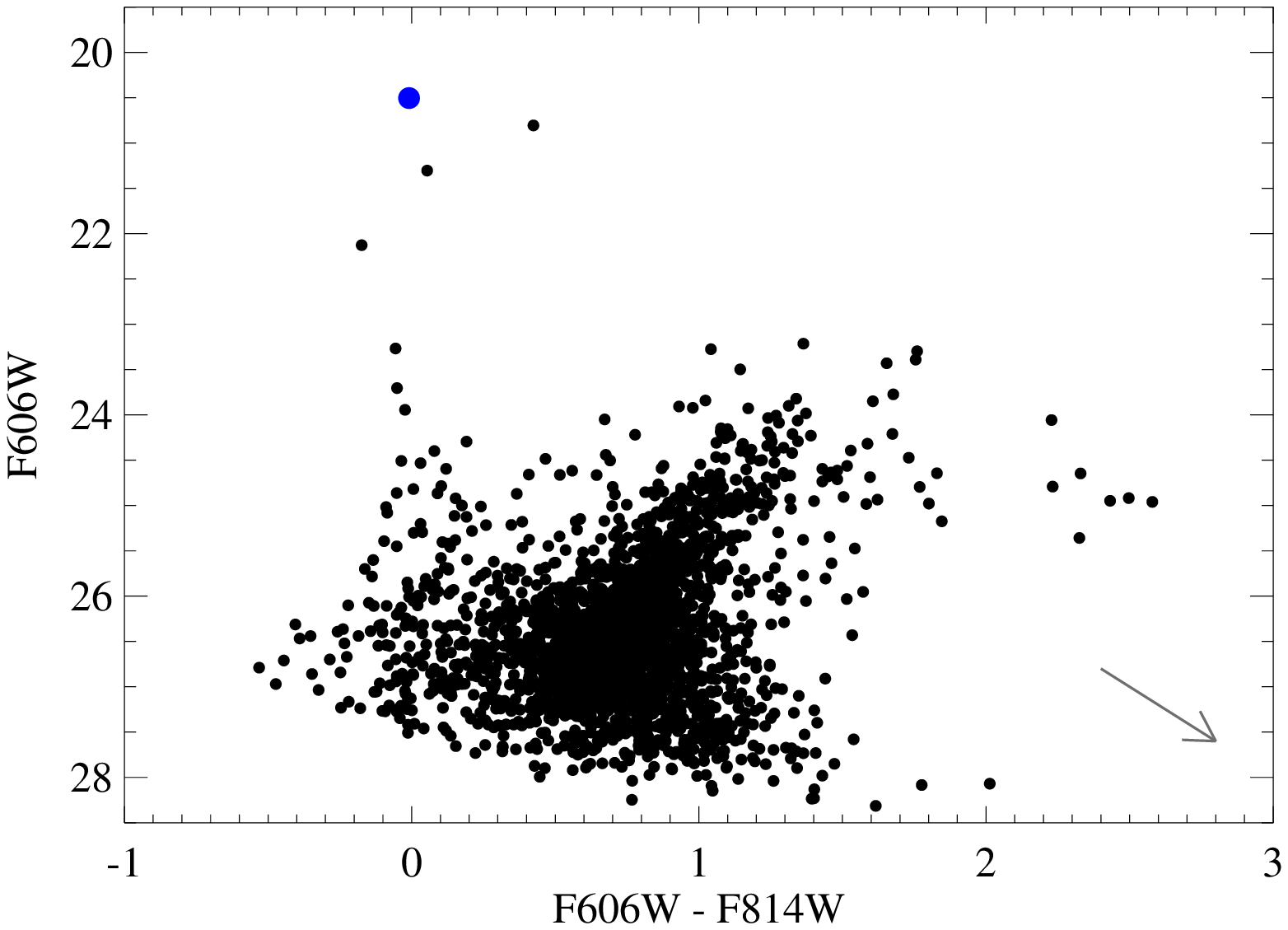} &
\includegraphics[width=0.5\linewidth,clip=true,trim=0cm 0cm 0cm 0cm]{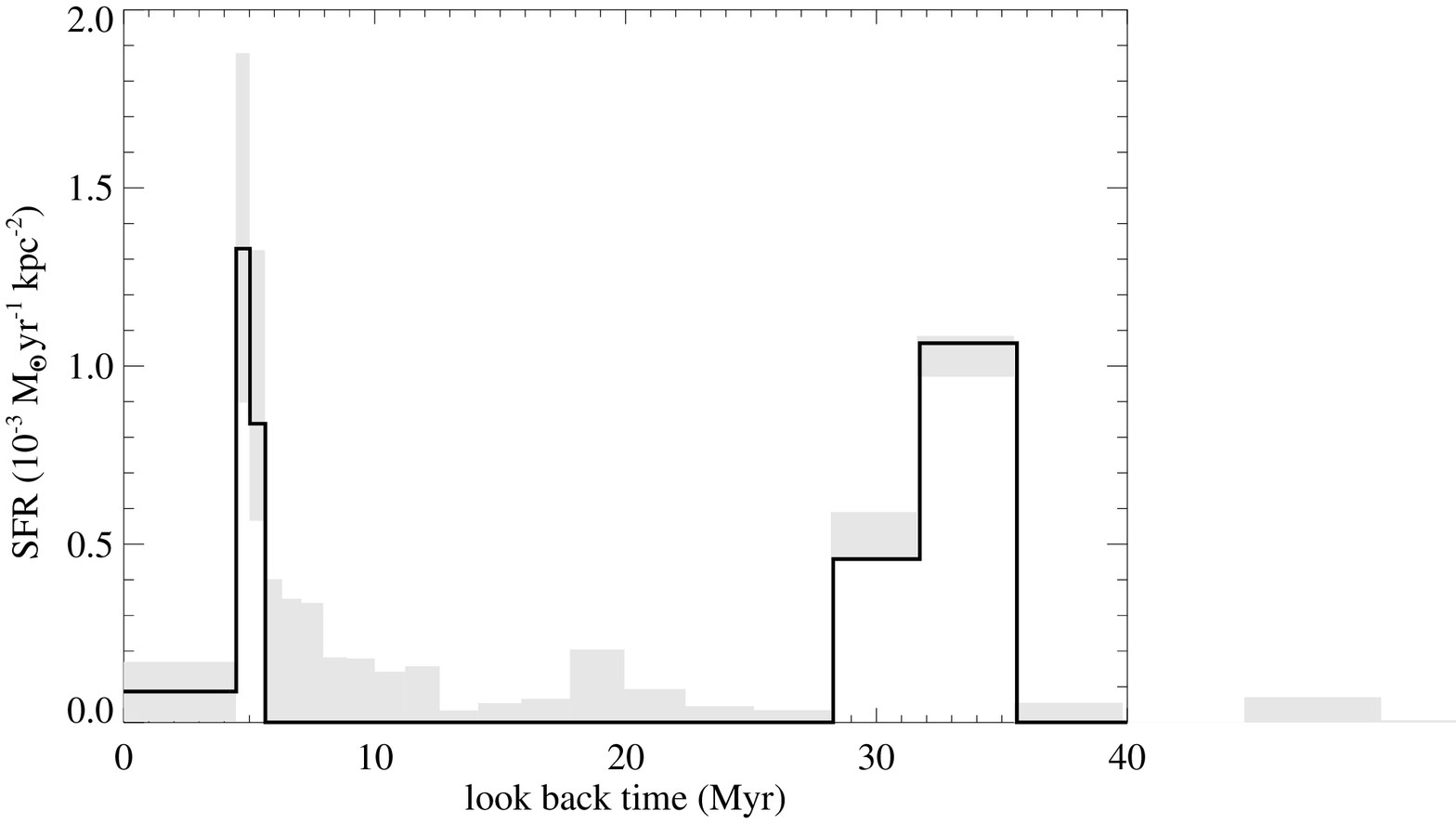} \\
\end{tabular}
\caption{{\it Left}: The CMD of stars within 50 pc of SN~2010da (the LBV companion is shown by the blue dot). The dark gray arrow shows the direction and magnitude of $E_{\rm F606W-F814W}=0.4$. {\it Right}: The recent star formation history of the young coeval stellar population. The gray shaded region indicates the uncertainty in the SFR in each age bin; the asymmetric uncertainties arise from the CMD methodology \citep[e.g., not simple counting statistics; see][for details]{Dolphin12,Dolphin13}. }
\label{figure_CMD}
\end{figure*}

We use the same technique here to determine the age of SN~2010da; for the remainder of this work, we define the age as the time since the formation of the progenitor binary. The CMD for all stars located within $\sim$50 pc of SN~2010da was modeled using the MATCH software package \citep{Dolphin02}, which produces a star formation history (SFH) that best describes the observed CMD. The fitting technique is described in detail in \cite{Williams+09}, and has been applied to a large number of galaxies as part of the ANGST survey \citep{Dalcanton+09}. Updates to the uncertainty estimation are described in \cite{Williams+13}. In addition to deriving the SFH at the location of SN~2010da, we also find the best-fitting mean foreground extinction to be $A_V=0.26$, equivalent to \nH $\sim5.6\times10^{20}$ cm$^{-2}$ \citep{Guver+09}. This value is consistent with the low absorbing columns found in Section~\ref{section_xray}.

We additionally include differential reddening in the model Hess diagrams using the MATCH \texttt{dAv} flag. We find a differential extinction (e.g., internal to NGC~300) of $\sim$0.4 mag at the location of SN~2010da, indicating that the majority of the dust surrounding the system prior to the 2010 outburst was indeed destroyed \citep[e.g., ][found a value of $A_V\sim12$ mag for the progenitor system]{Berger+10}. With the extinction values applied, we then run a series of 100 Monte Carlo tests. To assess the full combination of systematic errors due to model deficiencies as well as random errors due to the depth and size of the sample, realizations of the best-fitting model solution are fitted with the models shifted in bolometric magnitude and effective temperature \citep{Dolphin12,Dolphin13}. These shifts account for the uncertainties due to any potential systematic offsets between the data and models.

The CMD and resulting SFH, with the full error estimate, of the young coeval stellar population are shown in Figure~\ref{figure_CMD}. The SFH shows a sharp increase in the star formation rate (SFR) at $\lesssim$5 Myr, and a second peak at $\sim$35 Myr. An age of $\lesssim$5 Myr is consistent with zero-age main sequence mass of $>$20 \Msun. \cite{Foley+11} compared the $V$ and $I$ magnitudes of two SN impostor progenitors to stellar evolutionary tracks of \cite{Schaller+92} to determine the masses of the progenitor stars prior to outburst (their Figure~3); using the $F606W\approx V$ and $F814W \approx I$, we can constrain the mass of the LBV-like star to $\sim$15-25 \Msun. The optical companion is most likely an LBV and not a Be star; thus, the sibling population is likely the young $\lesssim$5 Myr one.

\section{Discussion and Conclusions}\label{section_discussion}	
	\subsection{The Origin of the X-ray Outbursts}
The new \Chandra X-ray observations provide the first evidence that the SN~2010da X-ray source, like the optical companion, may exhibit recurring outbursts in which the spectrum hardens at high luminosity. Furthermore, it is not known whether these X-ray outbursts occur simultaneously with brightening or erupting events from the LBV-like star, as the only simultaneous optical and X-ray observations are those obtained by the {\it Swift} XRT/UVOT during the initial 2010 outburst. There are two possible scenarios for the origin of the X-rays from SN~2010da: either the X-rays are produced by eruptions or mass ejection events from a massive star or massive binary, or they originate from accretion onto a compact object. Our observations alone cannot definitively rule out the stellar origin scenario. However, if the X-ray outbursts are from a massive star (or massive stellar binary), it would be extremely interesting given that such high X-ray luminosities have not yet been reported for any class of massive star \citep[even in the case of highly magnetized massive stars; see][ and references therein]{Petit+15}.

This is opposite of what is observed in black hole transients, in which the low-luminosity state is associated with a harder energy spectrum dominated by Comptonized emission that then transitions to a higher luminosity dominated by a softer accretion disk component \citep{McClintock+06}. Recently, some low-mass X-ray binaries with NS primaries have been observed to follow a similar low/hard-high/soft pattern \citep{MunozDarias+14}. If SN~2010da is indeed an HMXB, the observed X-ray emission may be related to the orbital period of the NS. No rapid variability was observed in any of our three \Chandra observations, indicating that the outburst duration is likely longer than the $\sim$60 ks \Chandra exposure times. Type~I outbursts are commonly observed in HMXBs, reach peak X-ray luminosities of a few 10$^{37}$ \lum for $\sim$20-30\% of the orbital period, and exhibit simultaneous brightening in the optical \citep{Reig08}. In known BeXBs, the orbital periods typically range from 20 to several hundred days \citep[][and references therein]{Reig11}, while HMXBs with supergiant companions can have orbital periods of 10 days or less \citep[][their Figure~2]{Reig11}. Assuming a NS mass of $\sim$1.4 \Msun, the observed X-ray luminosities can be explained by mass accretion rates of $\sim$2-15\% of the Eddington limit.

Whether the X-ray outbursts in SN~2010 are periodic cannot be determined from the available data. Further optical and X-ray monitoring of this system is needed to determine whether the X-ray outbursts are periodic, if they correspond to optical outbursts, and may help to distinguish between the stellar or compact object scenarios for the observed X-ray emission.

	\subsection{Binary Interactions as Drivers of LBV Outbursts}
LBVs represent a late stage of massive star evolution, lasting only $\sim$10$^4$ yr \citep{Humphreys+94,Smith+06,Harpaz+09}. Their frequent eruptions can often mimic SN explosions, leading to numerous LBVs with SN designations \citep{Smith+06,Davidson+12,Kochanek+12}. While the continually growing list of ``SN impostors'' is likely a heterogeneous mix of many varieties of transient sources, a large fraction of these events are likely related to eruptions of LBVs and other massive stars.

There is growing evidence that binary companions play a significant role in setting the conditions for giant LBV eruptions to occur \citep[e.g., through tidal interactions, see][and references therein]{Kashi+13}. For example, $\eta$~Car and P~Cygni \citep{Kashi10,Kashi+10}, SN~2009ip in NGC~7259 \citep{Kashi+13}, and the LBV + Wolf-Rayet binary in NGC~3432 \citep{Pastorello+10} are all systems where observed outbursts from the LBV can be explained by disturbances from a companion. While other impostors are known or suspected binary systems, the X-ray properties of these objects is largely unknown. If SN~2010da is a compact object accreting material from an LBV, the extreme gravitational environment of the NS would make SN~2010da an especially valuable target to study the effects of binary interactions and the evolution of massive close binaries. Alternatively, if the X-ray emission from SN~2010da originates from a massive star, it could provide a useful laboratory for studying recurring mass ejection events on a scale not observed in Galactic sources.

	\subsection{The Onset of the HMXB Phase?}
One extremely interesting possibility is that we are witnessing the birth of an HMXB. HMXBs dominate the X-ray emission from non-active star forming galaxies, and models of HMXB formation indicate that X-ray production first arises in these systems $\sim$5 Myr after the progenitor binary forms \citep{Linden+10, Postnov+14}. Although the typical timescale from compact object formation until X-ray production is expected to be $\sim$1-2 Myr \citep[e.g., see][and references therein]{Shty+05,Bodaghee+12}, other observations of young HMXBs, notably Cir~X-1 \citep{Heinz+13}, indicate that the onset of the HMXB phase may occur much more rapidly (within $\sim10^{4-5}$ yr). Stellar evolution models have been used to constrain the ages of dozens of HMXBs \citep[including those in NGC~300][]{Williams+13} and found them to have an average age of $\sim$40-55 Myr. In the LMC, HMXBs are found to only be associated with OB associations that are at least $\sim$10 Myr old \citep{Shty+05}. 

With an age (e.g., the time since the formation of the progenitor binary) of $\lesssim$5 Myr, SN~2010da represents an extremely young X-ray emitting HMXB. No X-ray emission was detected from the location of SN~2010da prior to the 2010 outburst, despite having been observed previously by both \XMM \citep{Carpano+05} and {\it ROSAT} \citep{Read+01}. It is possible that the 2010 outburst represented the initial onset of the HMXB phase in this system, possibly making SN~2010da the first massive evolved binary ever directly observed to enter the HMXB phase.

\section*{Acknowledgements}
We would like to thank the anonymous referee for suggestions that helped improve this manuscript. Support for this work was provided by the National Aeronautics and Space Administration through \Chandra Award Number G04-15088X issued by the \Chandra X-ray Observatory Center, which is operated by the Smithsonian Astrophysical Observatory for and on behalf of the National Aeronautics and Space Administration under contract NAS8-03060. TJG and PPP acknowledge support under NASA contract NAS8-03060. This research has made use of the NASA/IPAC Extragalactic Database (NED) which is operated by the Jet Propulsion Laboratory, California Institute of Technology, under contract with the National Aeronautics and Space Administration. This research has made use of the SIMBAD database, operated at CDS, Strasbourg, France.

\bibliographystyle{mnras}

\bsp	
\label{lastpage}
\end{document}